# KYOTO INSTITUTE
# OF
# ECONOMIC RESEARCH

Discussion Paper No.504

**A Bottleneck Principle
For
Techno-Metabolic Chains**

Almaz T. Mustafin

(Guest Scholar at KIER from the Space Research Institute,
Ministry of Science and Higher Education, Republic of Kazakhstan.
15 Shevchenko St., Almaty, 480100, Kazakhstan
E-mail: mustafin@mail.usa.com)

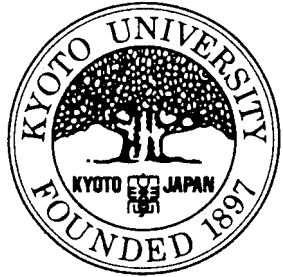

# KYOTO UNIVERSITY
# KYOTO, JAPAN

# A Bottleneck Principle
# for Techno-Metabolic Chains

Almaz T. Mustafin[*]

**Abstract**

A plausible mechanism of self-regulation in technological chains of the kind wherein input resources are converted into common final product through sequences of processing links having uniform kinetic properties is discussed. A bottleneck principle is derived according to which output of the overall chain is completely determined by its slowest link, if all the links are featured by satiation and weak outflow.

**Introduction**

Production function in its conventional formal notation $Y=f(R_1,\ldots,R_N)$ establishes the relationship between the volume of an output over certain period of time and the resources in use. In doing so the technology utilised in a given production system manifests itself as a converter of the input resources into a final product $P$. Output $Y$ has dimensionality "flow"; actually, it is the velocity of product yield: $Y = (\dot{P})_+$, where sign "+" implies the positive component of time derivative. The resources $R_1,\ldots,R_N$ involved in production function have dimensionality "stock".

The concept of production function is correct provided that the characteristic temporal scale of product dynamics far exceeds the duration of production cycle.

A path from the primary resources to the end product in a production system runs through, generally, ramified chain of elementary converters represented by technological installations, machines, tools, etc. In such a chain the product of one link serves as the resource for other. Once the notion of production function of an elementary link has been introduced, the following questions arise about the relation between macroscopic production function of the overall system under consideration and those of elementary links:

- Whether the velocity of production of a final good depends on parameters of all elementary links constituting the chain?

[*] Guest Scholar at KIER from the Space Research Institute, Ministry of Science and Higher Education, Republic of Kazakstan. E-mail: mustafin@mail.usa.com

- What are the conditions under which the reduction of description may take place typical for open non-linear systems economic ones, apparently, belong to?

In the present paper these questions are studied under rather broad assumptions about the mechanism of functioning of an elementary link. It is shown that production function of the overall chain is completely determined by velocity of resource processing in the slowest link.

**Linear chain**

Consider linear production chain constituted by $m$ links in which the initial resource $R=X_0$ is being processed into the final product $P=X_m$ through a sequence of intermediate products $X_i$ ($i=1,\ldots,m$-1) (Fig. 1).

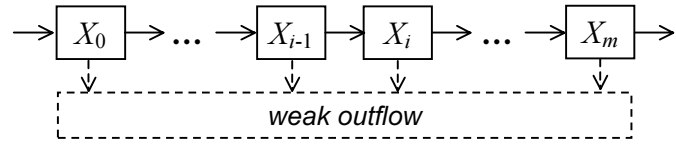

**Fig. 1.** Flow diagram of an open linear production chain.

We shall call such chains *techno-metabolic* ones, if each individual link is defined as an elementary input-output converter operating by the following model mechanism (Fig. 2).

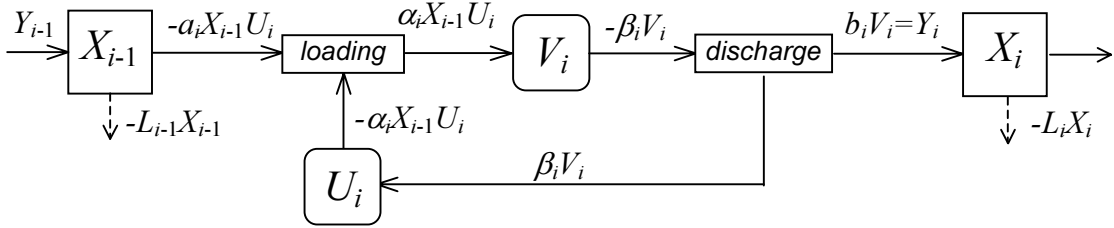

**Fig. 2.** Operation of an individual link in the linear chain. Expressions along the arrows pointing from inputs to outputs are the magnitudes of associated flows.

The intermediate product $X_{i-1}$ is transformed in the link $i$ into next intermediate product $X_i$ by means of processing units of fixed capacity. Those can be either in idle or loaded state. Let total number of units of a given type in the link under consideration be $W_i$, whereas $U_i$ and $V_i$ are the respective numbers of idle and loaded units at some instant of time.

In the first stage of processing an idle unit is loaded with the incoming resource. As this takes place, the velocity of resource binding is assumed to be proportional both to the volume of resource and the number of idle units with specific rate $a_i$. This is also true for the velocity of decrease of idle units but with specific rate $\alpha_i$.

In the second stage, in due course, the unit releases a ready-made product and returns to the idle state. The velocity of product release is proportional to the number of loaded units with specific rate $b_i$. The same is true for the velocity of increase of the number of idle units but with specific rate $\beta_i$.

Note that $a_i \neq \alpha_i$ and $b_i \neq \beta_i$, because dimensionalities of products and installations do not coincide.

In order for the techno-metabolic chain to operate in a steady state, the weak outflows of all intermediate products with specific rates $L_i$ ($i=1,\ldots,m-1$) should be allowed. Otherwise in some link a limitless accumulation of intermediate product will occur.

***Proposition 1.*** *If the total loading capacity of processing units in the elementary techno-metabolic link i is much less than the characteristic volume of the resource at hand, then production function of this link is expressed by the formula*

$$Y_i = b_i W_i X_{i-1} / (X_{i-1} + \beta_i / \alpha_i). \tag{1}$$

**Proof.** In line with the diagram in Fig. 1 the following set of balance equations may be written:

$$\begin{aligned} \dot{X}_{i-1} &= Y_{i-1} - L_{i-1} X_{i-1} - a_i U_i X_{i-1}, \\ \dot{U}_i &= \beta_i V_i - \alpha_i U_i X_{i-1}, \\ \dot{V}_i &= \alpha_i U_i X_{i-1} - \beta_i V_i, \\ Y_i &= \left(\dot{X}_i\right)_+ = b_i V_i. \end{aligned} \tag{2}$$

It is the fourth equation of set (2) that is a sought-for production function of the link. Its form is to be determined by solution of the closed subset of the first three equations containing no $X_i$. Combining the second and third equations we obtain an identity $U_i + V_i = \text{const} = W_i$ with respect to time. This enables variable $U_i$ to be excluded. In the remaining equations we may change to dimensionless variables of the order of unity using the scaling

$$\begin{aligned} x_{i-1} &= \alpha_i X_{i-1} / \beta_i, & v_i &= V_i / W_i, & \tau &= a_i W_i t, \\ y_{i-1} &= Y_{i-1} / (q_i b_i W_i), & y_i &= Y_i / (b_i W_i), \\ l_{i-1} &= L_{i-1} / (a_i W_i), & q_i &= a_i \beta_i / (b_i \alpha_i). \end{aligned}$$

Then set (2) takes the form

$$\dot{x}_{i-1} = y_{i-1} - (1 - v_i + l_{i-1})x_{i-1},$$
$$\varepsilon_i \dot{v}_i = x_{i-1} - (1 + x_{i-1})v_i, \qquad (3)$$
$$y_i = v_i,$$

where $\varepsilon_i = a_i W_i / \beta_i$ is a parameter representing ratio of the loading capacity of the all processing units $a_i W_i / \alpha_i$ to the characteristic volume of the available resource $\beta_i / \alpha_i$. Suppose, $\varepsilon_i \ll 1$. According to the *Tikhonov theorem* (e.g., [1]) an approximate solution of singularly perturbed set of equations (3), correct to the order of $O(1)$, may be found from the reduced set

$$\dot{x}_{i-1} = y_{i-1} - (1 - v_i + l_{i-1})x_{i-1},$$
$$x_{i-1} - (1 + x_{i-1})v_i = 0,$$
$$y_i = v_i.$$

The reduction procedure is correct because the steady state solution of the second equation from set (3) is stable. The sense of Tikhonov approximation is that variable $v_i$ is deemed "fast" as compared with variable $x_{i-1}$. Thereby on the time scale concerned, $\tau \gg \varepsilon$, variable $v_i$ stays in the quasi-steady state

$$v_i = x_{i-1}/(1 + x_{i-1}), \qquad (4)$$

tracking relatively smooth variations of $x_{i-1}$. In view of equation (4), "slow" equations for $x_{i-1}$ and $y_i$ take the form

$$\dot{x}_{i-1} = y_{i-1} - [1/(1 + x_{i-1}) + l_{i-1}]x_{i-1},$$
$$y_i = x_{i-1}/(1 + x_{i-1}). \qquad (5)$$

Turning back to the dimensional quantities in the second equation of set (5) we obtain formula (1). ■

The hyperbolic input-output relationship of the type (1) is widespread in biology: it describes the kinetics of enzyme reactions [2], growth rate of micro-organisms in nutrient solution [3], functional response of predator on prey population density [4], and the like. Thus, using the term "techno-metabolic chain" gains an informal content.

It should be noted, that dimensionless one-factor production function of an individual techno-metabolic link, $y=r/(1+r)$, has the following properties:

1) $y(0)=0$, i.e. output is impossible in the absence of the resource;

2) At low volumes of the available resource, $r \ll 1$, the output grows linearly. However as the volume of fed resource further increases, the output tends to its maximum. As this takes place, according to formula (4) all the available processing units are in the most use. When $r=1$, the units are half loaded;

3) $dy/dr = 1/(1+r)^2 > 0$, i.e. marginal product is always positive;

4) $d^2y/dr^2 = -2(1+r)^3 < 0$, i.e. there holds a law of diminishing marginal returns;

5) $E(r) = \lim_{\lambda \to 1}(\lambda/y(\lambda r))(dy(\lambda r)/d\lambda) = 1/(1+r) < 1$, i.e. production elasticity essentially depends on the input, while returns to scale may only diminish.

Fig. 3 depicts the production function, its marginal product and production elasticity.

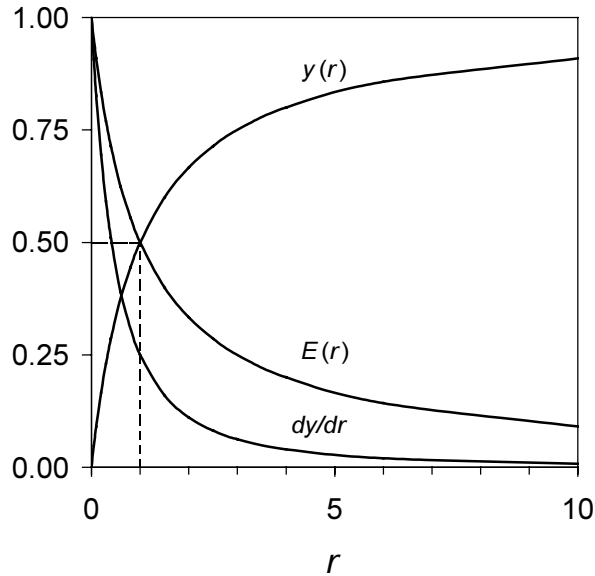

**Fig. 3.** Dimensionless production function of an individual techno-metabolic link ($y$), its marginal product ($dy/dr$) and production elasticity ($E$).

***Proposition 2.*** *Production function of the overall linear techno-metabolic chain of m links operating in a steady state is given by the formula*

$$Y = \min\{AR/(K+R), B\}, \qquad (6)$$

*where A is a maximal output of the first link, B is minimum amongst the greatest possible velocities of all links in the chain and K is a half-satiation constant. The parameters have the following expressions:*

$$A = b_1 W_1 \Big/ \prod_{j=2}^{m} q_j,$$
$$B = \min_{i=2,\ldots,m-1} \left\{ b_i W_i \Big/ \prod_{j=i+1}^{m} q_j, b_m W_m \right\}, \qquad (7)$$
$$K = \beta_1/\alpha_1, \quad q_j = a_j \beta_j / b_j \alpha_j, \; j = 1,\ldots,m.$$

*The quantities A and B are brought to the units of measurement of the end product flow.*

**Proof.** At steady state the volumes of all intermediate products remain constant:

$$R = X_0 = const,$$
$$\dot{X}_i = 0, \quad i = 1,\ldots,m-1.$$

Putting $\dot{x}_{i-1} = 0$ in set (5), we arrive at

$$y_{i-1} = x_{i-1}/(1+x_{i-1}) + l_{i-1} x_{i-1},$$
$$y_i = x_{i-1}/(1+x_{i-1}).$$

Eliminating $x_{i-1}$, we come to an equation relating production functions of two consecutive links:

$$y_i^2 - (y_{i-1} + 1 + l_i) y_i + y_{i-1} = 0,$$

which, by virtue of the above assumption that the specific rate of outflow of $(i-1)$th semi-product is reasonably weak ($l_{i-1} \ll 1$), has an approximate solution

$$y_i = \begin{cases} y_{i-1} + l_i/(y_{i-1} - 1), & y_{i-1} < 1, \\ 1 - l_i/(y_{i-1} - 1), & y_{i-1} > 1. \end{cases}$$

To a zeroth-order approximation in $l_{i-1}$, the following condition of stiff switching by velocity of processing takes place:

$$y_i = \min\{y_{i-1}, 1\},$$

which may be rewritten in dimensional form:

$$Y_i = \min\{Y_{i-1}/q_i, b_i W_i\}. \qquad (8)$$

The first term in braces is a normalised value of production function of ($i$–1)th link expressed in the units of velocity of $i$th link.

Successively applying the recurrent formula (8), we find a relationship between velocity of the last link and that of the first one, i.e. production function of the overall chain:

$$Y = Y_m = \min\left\{Y_1 \Big/ \prod_{j=2}^{m} q_j,\ \min_{i=2,\ldots,m-1}\left\{b_i W_i \Big/ \prod_{j=i+1}^{m} q_j, b_m W_m\right\}\right\}. \qquad (9)$$

Expressing $Y_1$ in terms of $R$ by formula (1), we arrive at the sought-for production function of the linear chain (6). ∎

Hence net velocity of production in the linear chain depends solely on the characteristics of the slowest link – "bottleneck". It is significant that close as the normalised processing velocities in the chain may be, switching among them remains stiff provided all the outflows are sufficiently weak.

Suppose the slowest link has number $s \in \{1,\ldots,m\}$. Then in the ($s$–1)th link an intermediate product $X_{s-1}$ will accumulate up to its steady-state volume $X_{s-1} = (Y_{s-1} - q_s b_s W_s)/L_{s-1}$. This peculiar kind of buffer arising immediately ahead the bottleneck completely eliminates an impact of the all preceding links on production velocity of $X_s$. As to the link $s$, the steady-state volume of its product has the order of $O(k_{s+1})$: $X_s = k_{s+1} Y_s / (q_{s+1} b_{s+1} W_{s+1} - Y_s)$, i.e. product $X_s$ does not accumulate since higher processing capacity of the subsequent links.

**Ramified chain**

Now consider such flow diagram that the final product of the chain number 0 results after the products of two other independent chains with the resources $R_1 = X_{10}$ and $R_2 = X_{20}$ being jointly processed (Fig. 4). $m_0$, $m_1$ и $m_2$ are the respective lengths of the three chains.

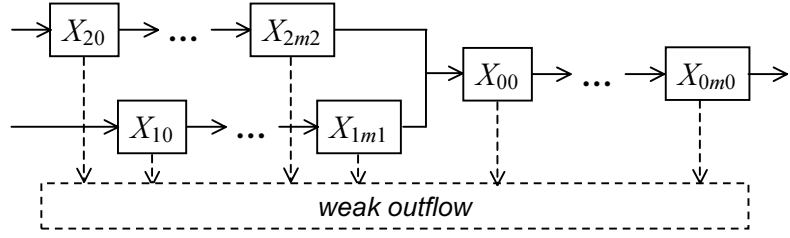

**Fig. 4.** Ramified techno-metabolic flow diagram, in which two independent linear chains are merged into one. First digit in the product's subscript stands for the number of a corresponding linear chain.

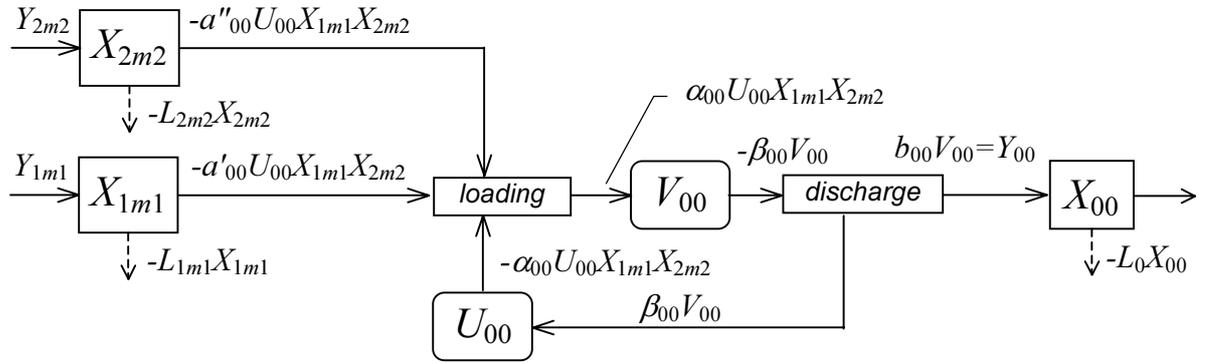

**Fig. 5.** Flow diagram of the Y-shaped link.

It turns out that the bottleneck principle is also true for ramified chain.

***Proposition 3.*** *At steady state production function of the techno-metabolic chain with two resources has a form of stiff switching by velocity-controlling resource:*

$$Y = \min\{A_1 R_1/(K_1 + R_1), A_2 R_2/(K_2 + R_2), B\}, \quad (10)$$

*where $A_1$ and $A_2$ are maximal outputs of the first links of the chains 1 and 2, B is minimum amongst the greatest possible velocities of all links in the all three chains, and $K_1$ and $K_2$ are half-satiation constants. The parameters have the following expressions:*

$$A_1 = b_{11}W_{11} \Big/ \Big(q'_{00} \prod_{j=1}^{m_0} q_{0j} \prod_{j=2}^{m_1} q_{1j}\Big),$$

$$A_2 = b_{21}W_{21} \Big/ \Big(q''_{00} \prod_{j=1}^{m_0} q_{0j} \prod_{j=2}^{m_2} q_{2j}\Big),$$

$$K_1 = \beta_{11}/\alpha_{11}, \quad K_2 = \beta_{21}/\alpha_{21},$$

$$B = \min\{B_0, B_1, B_2\},$$

$$B_0 = \min_{i=1,\ldots,m_0-1} \Big\{ b_{0i}W_{0i} \Big/ \prod_{j=i+1}^{m_0} q_{0j}, b_{0m_0}W_{0m_0} \Big\}, \quad (11)$$

$$B_1 = \min_{i=2,\ldots,m_1-1} \Big\{ b_{1i}W_{1i} \Big/ \prod_{j=i+1}^{m_1} q_{1j}, b_{1m_1}W_{1m_1} \Big\},$$

$$B_2 = \min_{i=2,\ldots,m_2-1} \Big\{ b_{2i}W_{2i} \Big/ \prod_{j=i+1}^{m_2} q_{2j}, b_{2m_2}W_{2m_2} \Big\},$$

$$q'_{00} = a'_{00}\beta_{00}/b_{00}\alpha_{00}, \quad q''_{00} = a''_{00}\beta_{00}/b_{00}\alpha_{00},$$

$$q_{nk} = a_{nk}\beta_{nk}/b_{nk}\alpha_{nk}, \quad n=0,1,2; k=1,\ldots,m_k$$

*(Here $B_0$, $B_1$ and $B_2$ are minima amongst the greatest possible velocities of the chains 0, 1 and 2). Controlling is that resource being most slowly processed.*

**Proof.** Consider the flow diagram for a joint of the three chains. Shown in Fig. 5 this is an Y-shaped link (having index "00") in which the intermediate products $X_{1m_1}$ and $X_{2m_2}$ originating from the two distinct resources $X_{10}$ и $X_{20}$ are jointly processed into a common intermediate product $X_{00}$. The relevant balance equations have the form

$$\dot{X}_{1m_1} = Y_{1m_1} - L_{1m_1}X_{1m_1} - a'_{00}U_{00}X_{1m_1}X_{2m_2} = 0,$$
$$\dot{X}_{2m_2} = Y_{2m_2} - L_{2m_2}X_{2m_2} - a''_{00}U_{00}X_{1m_1}X_{2m_2} = 0,$$
$$\dot{U}_{00} = \beta_{00}V_{00} - \alpha_{00}U_{00}X_{1m_1}X_{2m_2}, \quad (12)$$
$$\dot{V}_{00} = \alpha_{00}U_{00}X_{1m_1}X_{1m_2} - \beta_{00}V_{00},$$
$$Y_{00} = (\dot{X}_{00})_+ = b_{00}V_{00}.$$

It is essential to derive a relationship of production function $Y_{00}$ of the link under consideration and those $Y_{1m_1}$ and $Y_{2m_2}$ of the chains 1 and 2. Using the reasoning similar to that applied when obtaining formula (1), we find quasi-steady-state expressions for loaded and idle processing units:

$$U_{00} = W_{00}(\beta_{00}/\alpha_{00})/(X_{1m_1}X_{2m_2} + \beta_{00}/\alpha_{00}),$$
$$V_{00} = W_{00}X_{1m_1}X_{2m_2}/(X_{1m_1}X_{2m_2} + \beta_{00}/\alpha_{00}),$$

where $W_{00}=U_{00}+V_{00}=\text{const}$ is the total number of units in the link number 00. Substituting these formulae into set (12) and rearranging gives

$$Y_{1m_1}/a'_{00} - Y_{2m_2}/a''_{00} = L_{1m_1}X_{1m_1}/a'_{00} - L_{2m_2}X_{2m_2}/a''_{00},$$
$$Y_{00} = (Y_{1m_1} - L_{1m_1}X_{1m_1})/q'_{00} = (Y^{(2)}_{m_2} - L_{2m_2}X_{2m_2})/q''_{00}. \quad (13)$$

Introducing the designation $D = a'^{-1}_{00}Y_{1m_1} - a''^{-1}_{00}Y_{2m_2}$ we obtain the following expression from the first formula of set (13):

$$X_{1m_1} = a_{001}\left\{D + \left(D^2 + 4a'^{-1}_{00}a''^{-1}_{00}L_{1m_1}L_{2m_2}X_{1m_1}X_{2m_2}\right)^{1/2}\right\}/(2L_{1m_1}). \quad (14)$$

At reasonable weak outflows of the intermediate products the square root in formula (14) may be expanded in terms of the parameter $L_{1m_1}L_{2m_2}X_{1m_1}X_{2m_2}/(a'_{00}a''_{00}D^2) \ll 1$. Limiting the accuracy by zeroth order yields the following two different cases:
  1) If $D>0$ (chain 2 is slower than chain 1), then $X_{1m_1} = a'_{00}D/L_{1m_1}$;
  2) If $D<0$ (chain 1 is slower than chain 2), then $X_{1m_1} = -L_{2m_2}X_{1m_1}X_{2m_2}/(a''_{00}D)$, whence $X_{2m_2} = -a''_{00}D/L_{2m_2}$.

Substituting the obtained steady-state volumes of the intermediate products into the second equation of set (13) we get the production function of Y-shaped link which has the form of condition of stiff switching by velocity between the two chains:

$$Y_{00} = \min\{Y_{1m_1}/q'_{00}, Y_{2m_2}/q''_{00}\}. \quad (15)$$

It remains now to use previously established results on the linear chains. Substituting relationship (15) into recurrent equation (8) at $i=1$ and the latter – into formula (9), and using formulae for outputs $Y_{1m_1}$ and $Y_{2m_2}$ of the two independent chains in explicit form, we arrive at relationship (10). ∎

The proven proposition can be easily extended to an arbitrary number of the resources.

From the viewpoint of the structural approach, to construct a production function with switching by processing velocity using theoretical formula (10) calls for knowledge of the whole set of technological constants of the chain under study, which is not always possible. Because of this, in applications the functional approach seems to be more convenient instead, when formula (10) is treated as some phenomenological relationship, which parameters are to be empirically estimated.

## Discussion

It follows from the form of function (10) that switching over from one output-controlling resource to another may be possible only at unequal maximal normalised velocities of the primary links: $A_1 \neq A_2$. In the case of $A_2 > A_1$ the controlling factor change line lying in the resource plane $(R_1, R_2)$ is given by the equation

$$R_2 = A_1 K_2 R_1 / \left[ A_2 K_1 + (A_2 - A_1) R_1 \right] \qquad (16)$$

(see Fig. 6). Curve (16) has a horizontal asymptote $R_2 = A_1 K_2/(A_2 - A_1)$. Above the curve the resource $R_1$ controls overall output, whereas below the curve so does $R_2$. At $A_1 < B$ switching is possible only among $R_1$ and $R_2$. At $B < A_1$ a new domain $\{R_1 > K_1 B/(A_1 - B), R_2 > K_2 B/(A_2 - B)\}$ emerges in the resource plane where the output is controlled by some inner link of the chain.

Isoquants of production function (10) are L-shaped being broken in the controlling resource change line. Obviously, elasticity of substitution for the production function under consideration is equal to zero.

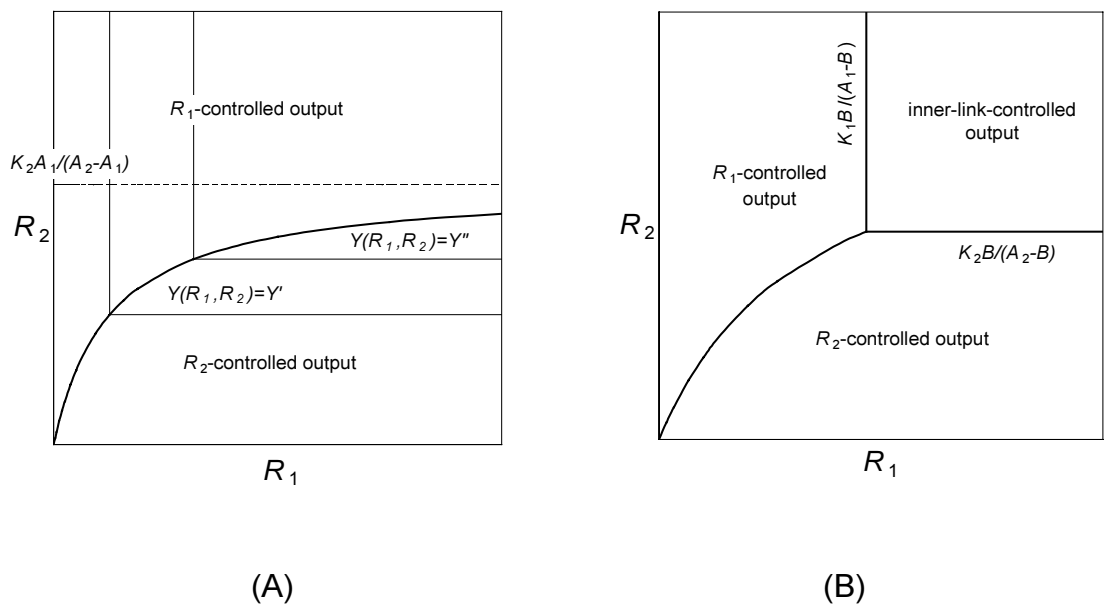

(A)          (B)

**Fig. 6.** The resource plane $(R_1, R_2)$ of the production function with stiff switching by resource processing velocity. Solid curve is the controlling resource change line. (A) Case of $A_2 > A_1$ and $B > A_1$. Output is controlled either by the first or second resource. $Y'$ and $Y''$ are the isoquants of the production function ($Y' < Y''$). (B) Case of $A_2 > A_1 > B$. At sufficiently high volumes of both resources output is controlled by some inner bottleneck.

It will be recalled from the theory of production functions that there has been known a function with zeroth elasticity of substitution somewhat resembling equation (10). The case in point is the *Leontief production function* with constant proportions of the inputs having the form of condition of stiff switching by volume of controlling resource:

$$Y_L = \gamma \min\{R_1/\rho_1, R_2/\rho_2\}, \qquad (17)$$

where $\gamma$, $\rho_1$ and $\rho_2$ are positive parameters. The controlling factor change line for function (17) is a straight line $R_2 = \rho_2 R_1/\rho_1$ (Fig. 7).

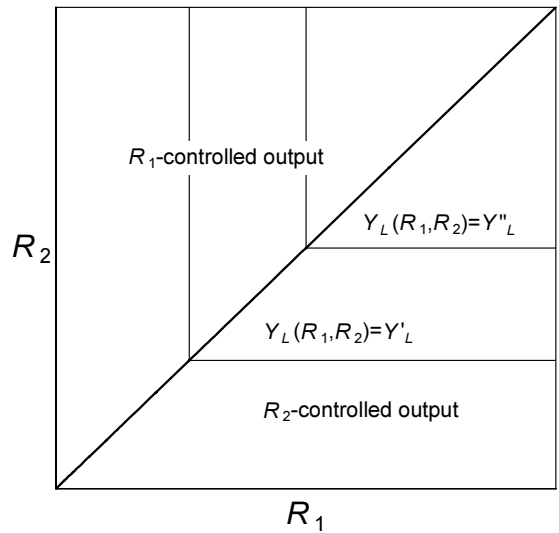

**Fig. 7.** The resource plane of the Leontief production function. The controlling resource change line looks like a straight line. $Y'$ and $Y''$ are isoquants of the production function ($Y' < Y''$).

It is our belief that the principle of switching by resource volume results from the stated above principle of switching by processing velocity in a bottleneck. Indeed, piecewise linear approximation of equation (10) gives

$$Y = \min\{A_1 R_1/K_1, A_2 R_2/K_2, B'\},$$
$$B' = \min\{A_1, A_2, B\},$$

that coincides with equation (17) at $R_1 < K_1 B'/A_1$ and $R_2 < K_2 B'/A_2$.

The emergence of a mechanism of stiff switching by velocity arises from the following properties of production function of an individual techno-metabolic link: satiation of output with increasing input and existence of weak outflow of intermediate products. This principle of self-regulation is important both for understanding the operation of complex technological systems and for their modelling. The fact that the overall output depends solely upon the production function of the bottleneck enables the system to avoid information overload and to simplify the problem of control by acting only on the controlling link. When describing the process of production in terms of formalism of techno-metabolic chains one need only to have knowledge of the parameters having an impact on the slowest stage of resource processing.

## Conclusion

In the present paper, we considered a plausible mechanism responsible for the reduction of information required for controlling the output in a complex chain of resource processing. It lies in the fact that production function of the overall chain is completely determined by velocity of resource processing in the slowest link. Stiff switching by velocity of processing is brought about by the property of satiation of production function of an individual link and by the existence of weak outflows of intermediate products. Ultimately, it is due to non-linearity and openness of the technological system. In turn these features are known to be critical requirements of evolutionary adaptability of any self-organising system.

The support of this work from the Monbusho International Research Programme (Joint Research, Subject No. 09044028) led by Prof. Tsuneo Tsukatani should be acknowledged.

# KIER ディスカッション・ペーパー・シリーズ

(平成10年度)

No.490 Koichi Futagami and Akihisa Shibata, "Welfare Effects of Bubbles in an Endogenous Growth Model," January 1999.

No.491 Akira Okada and Arno Riedl, "Inefficiency and Social Exclusion in a Coalition Formation Game: Experimental Evidence," January 1999.

No.492 A.K. Kantarbayeva, U.E. Shukeyev and T. Tsukatani, "Enterpreneurial Environment in Kazakhstan: Elimination of the Existing Barriers," February 1999.

No.493 Masahisa Fujita and Tomoya Mori, "A Flying Geese Model of Economic Development and Integration: Evolution of International Economy a la East Asia," February 1999.

No.494 Satoshi Mizobata, "Russia's Financial Crisis and Banking Sector Reorganization," March 1999.

No.495 Akira Okada, "A Cooperative Game Analysis of $CO_2$ Emission Permits Trading: Evaluating Initial Allocation Rules," March 1999.

No.496 Kimio Morimune and Mitsuru Nakagawa, "Power Comparisons of the Discontinuous Trend Unit Root Tests," March 1999.

(和文)

No.9605 朱金海、戦後日本産業政策与上海経済、1996年8月。

No.9701 坂井昭夫、国際的相互依存論とは何か？―国際政治経済学 サーベイの一幕―、1997年6月。

No.9702 藤木裕、農作物貿易システムの変化とコメの関税化・国内自由化、1998年2月。

No.9801 岩本康志、財投債と財投機関債、1998年5月。

No.9802 岩本康志、2020年の労働力人口、1998年7月。

No.9803 藤田昌久、久武昌人、日本と東アジアにおける地域経済システムの変容 新しい空間経済学の視点からの分析、1998年9月。

No.9804 坂井昭夫、国際公共財としての通貨システム、1999年2月

(平成11年度)

No.497 Akira Okada and Arno Riedl, "When Culture does not Matter: Experimental Evidence from Coalition Formation Ultimatum Games in Austria and Japan," April 1999.

No.498 Futoshi K. Yamauchi and Guillermo Abdel Musik, "Learning, Catch-up and the Spillover of Information: Evidence from Foreign Automobile Makers in Mexico," May 1999.

No.499 Harrison Cheng, "Efficiency and Agency Costs in Repeated Team Agency," June 1999

No.500 Kenn Ariga, Yasushi Ohkusa and Kiyohiko G. Nishimura, "Determinants of Individual Firm Mark-up in Japan : Market Concentration, Market Share and FTC's Regulations," July 1999

No.501 Hitoshi Matsushima, "The Role of Mobility among Regions in Coordination," July 1999

No.502 Jess Benhabib and Kazuo Nishimura, "Indeterminacy Under Constant Returns to Scale in Multisector Economies," December 1998

No.503 Harrison Cheng, "Optimal Collusive Oligopolies with Moral Hazard," August 1999

(和文)
No.9901 坂井昭夫、米国経済の変容とニューエコノミー論、1999年8月